\documentclass[10pt,conference]{IEEEtran}

\IEEEoverridecommandlockouts

\usepackage[T1]{fontenc}
\usepackage[utf8]{inputenc}
\usepackage{graphicx}
\usepackage{amsmath,amssymb}
\usepackage{booktabs}
\usepackage{multirow}
\usepackage{xcolor}
\usepackage{comment}
\usepackage[activate]{microtype}
\usepackage[numbers,sort&compress]{natbib}
\usepackage[hidelinks]{hyperref}
\usepackage[capitalize,noabbrev]{cleveref}

\title{Why Tell an Audio Language Model What It Can Already Hear?}

\author{
\IEEEauthorblockN{
Iosif Tsangko\IEEEauthorrefmark{1}\IEEEauthorrefmark{2},
Andreas Triantafyllopoulos\IEEEauthorrefmark{1}\IEEEauthorrefmark{2},
Björn W. Schuller\IEEEauthorrefmark{1}\IEEEauthorrefmark{2}\IEEEauthorrefmark{3}\IEEEauthorrefmark{4}
\thanks{This work was partially funded from the DFG's Reinhart Koselleck project No.\ 442218748 (AUDI0NOMOUS) and the EU H2020 project No.\ 101135556 (INDUX-R).}
}

\IEEEauthorblockA{\IEEEauthorrefmark{1}
CHI -- Chair of Health Informatics, TUM University Hospital, Munich, Germany}

\IEEEauthorblockA{\IEEEauthorrefmark{2}
MCML -- Munich Center for Machine Learning, Munich, Germany}

\IEEEauthorblockA{\IEEEauthorrefmark{3}
MDSI -- Munich Data Science Institute, Munich, Germany}

\IEEEauthorblockA{\IEEEauthorrefmark{4}
GLAM -- Group on Language, Audio, \& Music, Imperial College London, UK}
}

\begin{document}

\maketitle
\begingroup
\renewcommand{\thefootnote}{}
\footnotetext{\copyright\ 2026 IEEE.
Personal use of this material is permitted.
Permission from IEEE must be obtained for all
other uses, in any current or future media,
including reprinting/republishing this material
for advertising or promotional purposes,
creating new collective works, for resale or
redistribution to servers or lists, or reuse
of any copyrighted component of this work
in other works.}
\endgroup
\begin{abstract}
Reasoning over interpretable intermediate evidence has improved large language model (LLM) performance, but it remains unclear what reasoning should look like for audio language models (ALMs), where this evidence is continuous and acoustic. We study speech emotion recognition (SER) and propose to use interpretable acoustic concepts as such intermediate evidence. Before training ALMs to generate these concepts, we test their effectiveness in the oracle setting: six concept tokens (energy, pitch, brightness, dynamics, formants, and voice quality) derived label-free from eGeMAPS and appended to the prompt, with the audio input unchanged. Unlike prior work supplying acoustic descriptions to text-only LLMs, our ALMs also accept audio as input, letting us probe how textual and acoustic channels are integrated. Across FAU-Aibo and IEMOCAP, aligned tokens consistently improve unweighted average recall. Token-only interventions (shuffle, conflict, graded corruption) show that the gain is largely content-driven and degrades monotonically as token quality decreases, while perturbed-token conditions remain close to the audio-only baseline. This suggests that ALMs use symbolic acoustic evidence, but are not dominated by it, motivating models that generate such concepts themselves.
\end{abstract}


\section{Introduction}

Instruction-following \emph{audio language models} (ALMs) couple an audio front-end with an autoregressive language model, enabling prompt-based inference and natural language rationales across speech and audio tasks \citep{Goel25-AFA,Triantafyllopoulos25-CAF}. In text-based LLMs, gains increasingly come from interpretable intermediate reasoning; its form for ALMs remains unclear because task-relevant evidence is continuous and only partly expressible in language~\citep{Jacovi20-TFI,DeYoung20-ERB}. 
This is particularly important for speech emotion recognition (SER), which emerging regulatory frameworks increasingly consider a ``high-risk'' application requiring interpretable automated decisions~\citep{AIAct}.

SER is a core topic in computational paralinguistics, where models infer affective states from vocal cues such as prosody, spectral shape, and voice quality~\citep{Schuller14-CEP}. Over the last years, the field has moved from task specific pipelines towards foundation style representations, including self-supervised speech encoders~\citep{Baevski20-WAF,Hsu21-HSM,Chen22-WLP} and cross modal audio text pretraining \citep{Elizalde23-CLA, Schuller26-ACH}. With modern ALMs, this shift also enables prompt-based SER: one can prompt a model with an utterance and request a discrete label, optionally accompanied by a natural language justification.

Classic SER research relied on interpretable, handcrafted acoustic descriptors~\cite{Eyben16-GMA,Eyben10-OSM}, linked to emotional expression over decades. These descriptors are natural candidates for audio reasoning: an ALM should plausibly identify such properties and combine them with deeper signal representations. The long-term goal is therefore ALMs that \emph{generate} concept-level intermediate steps themselves. As a precursor, we test the \emph{oracle setting}: do contemporary ALMs benefit when correctly extracted concepts are provided in context, and are these gains driven by acoustic content rather than the mere addition of a structured token block? Crucially, unlike recent work supplying textual acoustic descriptions to \emph{text-only} LLMs~\cite{Santoso24-LEA,Wu25-BSL,Li25-RRR}, our models also receive the audio; the concept tokens are partly redundant given the audio input, and any benefit points to a gap between what ALMs \emph{perceive} and use.

In this work, we introduce categorical \emph{concept tokens} by binning interpretable acoustic descriptors extracted from the target utterance. We build on the \emph{eGeMAPS} feature set~\citep{Eyben16-GMA} by summarising features into six prosodic and voice-related categories: energy, pitch, dynamics, brightness, formants, and voice quality. These binned concept tokens are appended to the ALM prompt as structured auxiliary cues. We then probe audio-grounded reasoning by perturbing the concept tokens while holding the audio input fixed. Concretely, we test aligned tokens, shuffled tokens, deliberately conflicting tokens, and graded token corruption. We operationalise integration behaviourally: a model that uses the concept \emph{content} should benefit from aligned tokens, degrade monotonically as they are corrupted, and remain anchored to the audio when they mislead. The extent to which ALM predictions respond to these token-only interventions allows us to estimate whether the models integrate symbolic cues with the audio, ignore the cue channel, or over-follow misleading prompt information.

\begin{table*}[t]
\centering
\caption{Concept tokens derived from \emph{eGeMAPS} mean descriptors. We use \(\mu(\cdot)\) to denote the file-level mean of the corresponding \emph{eGeMAPS} dimension.}
\label{tab:concept_tokens}
\footnotesize
\setlength{\tabcolsep}{3pt}
\begin{tabular}{l p{0.48\textwidth} p{0.34\textwidth}}
\toprule
\textbf{Concept} & \textbf{\emph{eGeMAPS} mean descriptor(s) (compact)} & \textbf{Interpretation / tokenisation} \\
\midrule
ENERGY &
$\mu(Loudness)$ &
Overall intensity proxy (loudness). Discretised into 5 quantile bins: \texttt{VERY\_LOW}--\texttt{VERY\_HIGH}. \\

PITCH &
$\mu(F0)$ &
Fundamental frequency (F0) in semitone scale. Discretised into 5 quantile bins; on IEMOCAP, bin edges are computed separately per gender. FAU-Aibo contains pre-pubertal children, so gender-specific pitch/formant binning is not applied. \\

BRIGHTNESS &
$\mu(AR)$, $\mu(HI)$, $\mu(Slope_{0Hz-500Hz})$, $\mu(Slope_{500Hz-1500Hz})$ &
Spectral tilt / spectral energy distribution proxies (brighter vs darker timbre). Discretised into 5 quantile bins: \texttt{VERY\_DARK}--\texttt{VERY\_BRIGHT}. \\

DYNAMICS &
$\mu(Flux)$ &
Spectral change rate over time (more/less dynamic signal). Discretised into 5 quantile bins: \texttt{VERY\_STATIC}--\texttt{VERY\_DYNAMIC}. \\

FORMANTS &
$\mu(F1_f)$, $\mu(F2_f)$, $\mu(F3_f)$ &
Resonant structure of the vocal tract (formant frequencies). Discretised into 5 quantile bins; on IEMOCAP, bin edges are computed separately per gender. \\

VOICE\_QUALITY &
$\mu(Jitter)$, $\mu(Shimmer)$, $\mu(HNR)$, $\mu(H1\!-\!H2)$, $\mu(H1\!-\!A3)$ (and $\mu(F0)$ as auxiliary cue) &
Heuristic categorical label from perturbation/noise and harmonic-structure cues: \texttt{MODAL}/\texttt{BREATHY}/\texttt{PRESSED}/\texttt{ROUGH}/\texttt{CREAKY} (optionally \texttt{UNVOICED}). \\
\bottomrule
\end{tabular}
\end{table*}

\section{Related Work}

Recent ALM evaluation is increasingly benchmark-driven, with broad task coverage and unified scoring protocols proposed in AudioBench~\citep{wang-etal-2025-audiobench}, AIR-Bench~\citep{yang-etal-2024-air}, and VocalBench~\citep{liu2025vocalbench}. Beyond aggregate benchmark scores, recent work focuses on reliability issues, such as hallucinations and errors on detecting individual events~\citep{kuan24}, with mitigation strategies based on synthesising negative samples~\citep{kuan25}. 
Reasoning-oriented assessment for ALMs has also been formalised via multi-hop reasoning benchmarks~\citep{yang25g} and temporal reasoning with confidence estimation via perturbation-invariance~\citep{bhattacharya25b}, aligning with our intervention-based consistency analysis. 
Finally, evidence that (largely) frozen LLM backbones can perceive paralinguistic attributes through learnt speech interfaces~\citep{kang25} supports our use of concept/paralinguistic cues for improving the performance and reliability of ALMs.

\textbf{Audio foundation models for SER and computational paralinguistics.}
\emph{Computational paralinguistics} links speech to speaker states and traits, with SER as one of its central tasks \citep{Schuller14-CEP}. Earlier SER pipelines relied on expert-designed acoustic descriptors, typically extracted with toolkits such as \emph{openSMILE} \citep{Eyben10-OSM}. Widely used feature sets include \emph{eGeMAPS} \citep{Eyben16-GMA} (among others), which provide an interpretable interface for affective analysis and benchmarking \citep{Schuller09-IEC,Wagner23-DTT}. More recently, self-supervised speech encoders (wav2vec~2.0, HuBERT, WavLM) have become standard backbones for downstream paralinguistic prediction \citep{Baevski20-WAF,Hsu21-HSM,Chen22-WLP,Wagner23-DTT}.
In parallel, \emph{contrastive audio-text pretraining} (e.g., CLAP) aligns audio and text in a shared embedding space, enabling zero-shot matching \citep{Elizalde23-CLA}. For computational paralinguistics in particular, ParaCLAP leverages feature-derived prompting to form text queries, with SmoothCLAP additionally introducing soft-target supervision to better reflect graded affective similarity \citep{Jing24-PTA,Jing26-SSE}. While these contrastive encoders provide strong retrieval-style baselines, they are not designed to consume structured, per-utterance auxiliary cues.
\emph{Generative} ALMs instead couple an audio encoder with an instruction-following autoregressive LLM, enabling audio-conditioned text generation and structured outputs. Representative open-source models include Qwen2-Audio \citep{Xu24-Q2A}, Qwen2.5-Omni \citep{Xu25-QOT}, and Audio Flamingo~3 \citep{Goel25-AFA}. Very recent work has tried to explicitly target multi-step \emph{audio reasoning} (e.g., Audio-Reasoner and ECHO), motivating evaluation protocols that go beyond aggregate accuracy and probe how models use intermediate information \citep{Chen25-ARL,Gao26-ECH}. Finally, although explainability for SER has advanced (often via feature attribution and post-hoc analyses), such explanations do not by themselves establish \emph{faithful} use of provided cues, motivating our intervention-based tests \citep{Nfissi24-UHF, Alican25-IAE}.

\textbf{Faithfulness and intervention based evaluation of explanations.}
A broad interpretability literature has shown that explanations can be plausible without being faithful to the underlying decision process~\citep{Jacovi20-TFI,DeYoung20-ERB}. Common evaluation strategies include perturbation-based tests, in which the basic assumption is that removing or altering purported cues should change the prediction~\citep{Jain19-ANE,Adebayo18-SCS, Alican25-IAE}. These ideas have recently been revisited in the context of large models, where internal interventions can reveal causal structure beyond post hoc rationales~\citep{Jain25-IFI}. In the audio-language setting, neuron level studies have provided complementary evidence that targeted interventions can selectively modulate emotion-related behaviour \citep{Zhao26-DCV}. Our work follows the intervention principle, but focuses on \emph{explicit} auxiliary cues that are visible to the language model at inference time, enabling controlled tests without modifying model weights or internal activations. Our concept tokens can also be viewed as a test-time, textual analogue of concept bottleneck models, which expose interpretable concepts as an intervenable intermediate interface~\cite{Koh20-CBM}.

\textbf{Audio-grounded reasoning in ALMs.} ALMs support flexible prompting, facilitating the \emph{inclusion of auxiliary context} and allowing for open-ended outputs in natural language. This makes them attractive for speech tasks, where one may ask for a discrete label, a rationale, or a structured report~\citep{Peng25-ASO,Su25-ALM}. A complementary line supplies textual descriptions of acoustic properties to \emph{text-only} LLMs for emotion recognition, for example through acoustic feature descriptors in annotation prompts~\cite{Santoso24-LEA}, natural-language renderings of vocal nuances~\cite{Wu25-BSL}, or emotion-specific prompts that include acoustic knowledge~\cite{Li25-RRR}. In those settings, the descriptors compensate for a missing audio channel; in our setting, the ALM already receives the waveform, turning the question into one of \emph{channel integration}. However, the extent to which ALMs use the additional information provided through the prompt, rather than relying primarily on the audio representation, has not been widely quantified. Recent work on audio reasoning highlights that \emph{reasoning} should not be assumed from fluent outputs:~\citep{Fan25-CESAR} shows that prompting audio LLMs to produce longer chain-of-thought can reduce accuracy and even collapse under test-time scaling, motivating behavioural probes that test how predictions respond to controlled changes in available information. We address this gap by supplying interpretable \emph{concept tokens} grounded in established paralinguistic descriptors and by evaluating audio-grounded reasoning behaviour through controlled, token-only interventions.

\begin{figure}[t]
  \centering  \includegraphics[width=1.15\linewidth]{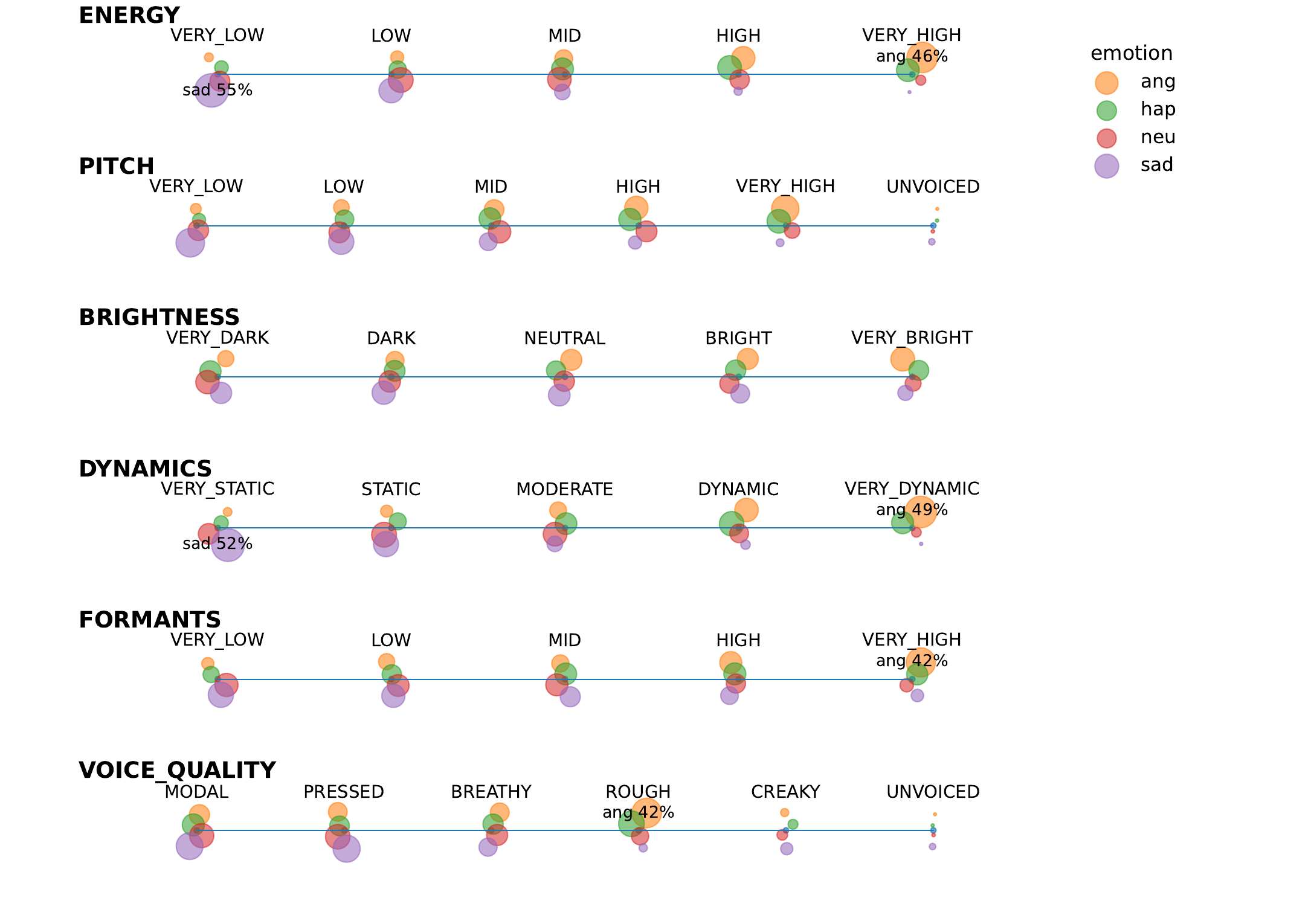}
    \caption{IEMOCAP (4-way) concept-token distributions across emotions. For each prosodic concept (row), bubbles show the within-emotion frequency (\%) of each discrete token value; bubble area is proportional to frequency. Small jitter is applied only for visual separation.}
  \label{fig:dotstrips}
\end{figure}

\section{Methodology}

% -------------------- AIBO5 --------------------
\begin{table}[t]
\centering
\setlength{\tabcolsep}{4pt}
\renewcommand{\arraystretch}{1.05}
\small
\caption{AIBO5 UAR without concept tokens ($UAR^{-}$) and with appended concept tokens ($UAR^{+}$). $\Delta = UAR^{+}-UAR^{-}$.}
\label{tab:aibo5}
\begin{tabular}{@{}l ccc ccc@{}}
\toprule
\multirow{2}{*}{\textbf{Model}} &
\multicolumn{3}{c}{\textbf{dev}} &
\multicolumn{3}{c}{\textbf{test}} \\
\cmidrule(lr){2-4}\cmidrule(lr){5-7}
& \textit{$UAR^{-}$} & \textit{$UAR^{+}$} & \textit{$\Delta$}
& \textit{$UAR^{-}$} & \textit{$UAR^{+}$} & \textit{$\Delta$} \\
\midrule
\multicolumn{7}{@{}l}{\textbf{Zero-shot ALMs}} \\
qw2-audio & .209 & .240 & \textcolor{green!50!black}{+.031} & .206 & .234 & \textcolor{green!50!black}{+.028} \\
qw-omni   & \textbf{.262} & \textbf{.279} & \textcolor{green!50!black}{+.017} & .230 & .240 & \textcolor{green!50!black}{+.010} \\
AF3       & .259 & .268 & \textcolor{green!50!black}{+.009} & \textbf{.253} & \textbf{.269} & \textcolor{green!50!black}{+.016} \\
\midrule
\multicolumn{7}{@{}l}{\textbf{Fine-tuned ALMs}} \\
ft-qw-omni & .260 & .263 & \textcolor{green!50!black}{+.003} &  -- &  -- & -- \\
ft-AF3     & .248 & .268 & \textcolor{green!50!black}{+.020} &  -- & -- & -- \\
\bottomrule
\end{tabular}
\end{table}

\textbf{Datasets}. We evaluate on two established speech emotion recognition benchmarks that differ markedly in language, speaker demographics, and recording conditions.

\underline{\textbf{FAU-Aibo (AIBO5).}} The FAU-Aibo Emotion Corpus is a German dataset of spontaneous children's speech (ages 6--10\,years) recorded in a naturalistic human--robot interaction setting~\citep{Batliner08-FAU}. It also served as the dataset for the INTERSPEECH 2009 Emotion Challenge~\citep{Schuller09-IEC}. Following common practice, we use the standard mapping from the original fine-grained annotations to five classes (\textit{angry}, \textit{emphatic}, \textit{neutral}, \textit{positive}, \textit{rest}). Since FAU-Aibo does not provide an official development split, we use the validation split adopted in prior work~\cite{Triantafyllopoulos26-CSP} for broad model sweeps and rapid prompt/token ablations, and reserve the official test split for the final comparison of the strongest ALM backbones.

\underline{\textbf{IEMOCAP (4-way).}} IEMOCAP is an English corpus of acted dyadic interactions with scripted and improvised speech~\citep{Busso08-IEM}. To align with prior work and to ensure sufficient support per class, we adopt the widely used 4-way setup (\textit{angry}, \textit{happy+excited}, \textit{neutral}, \textit{sad}), where \textit{excited} is merged into \textit{happy} and the remaining categories are discarded~\citep{Antoniou23-DAE}. This yields 5{,}531 utterances in total. \cref{fig:dotstrips} visualises the per-class distributions of the mapped concept tokens for this dataset. We also exclude other standard datasets, such as MSP-Podcast, from our evaluation as they are part of the training data for some of the ALMs considered, which would confound generalisation.

% -------------------- IEMOCAP --------------------
\begin{table}[b]
\centering
\setlength{\tabcolsep}{5pt}
\renewcommand{\arraystretch}{1.05}
\small
\caption{IEMOCAP (full) UAR with/without appended concept tokens. Contrastive zero-shot references are reproduced from~\citep{Jing26-SSE}.}
\label{tab:iemocap}
\begin{tabular}{@{}l ccc@{}}
\toprule
\textbf{Model} & \textit{$UAR^{-}$} & \textit{$UAR^{+}$} & \textit{$\Delta$} \\
\midrule
\multicolumn{4}{@{}l}{\textbf{Audio language models}} \\
AF3     & \textbf{.754} & \textbf{.776} & \textcolor{green!50!black}{+.022} \\
qw-omni & .541 & .582 & \textcolor{green!50!black}{+.041} \\
\midrule
\multicolumn{4}{@{}l}{\textbf{Zero-shot contrastive baselines (no tokens)}} \\
CLAP       & \multicolumn{1}{c}{.353} \\
Pengi      & \multicolumn{1}{c}{.345} \\
ParaCLAP   & \multicolumn{1}{c}{.600} \\
SmoothCLAP & \multicolumn{1}{c}{.606} \\
\bottomrule
\end{tabular}
\end{table}

\textbf{Evaluation protocol.} All datasets are evaluated using UAR to account for class imbalance. We report $UAR^{-}$ for the audio-only prompt (no appended concept tokens) and $UAR^{+}$ for the audio+concept-tokens prompt; $\Delta = UAR^{+}-UAR^{-}$. On AIBO5, we first run a multi-model sweep on the development set to characterise ALM behaviour under prompting and token augmentation. We then report test performance for the base backbones (qw-omni and AF3). On IEMOCAP, we report UAR for both qw-omni and AF3, and conduct extended intervention analyses on AF3, the most reliable backbone, to probe information faithfulness under controlled perturbations of the appended concept tokens. For all IEMOCAP comparisons, we report 95\% confidence intervals from a speaker-clustered bootstrap ($10$ speakers, $10^4$ resamples) and test paired differences with an exact McNemar test on per-utterance correctness.

\textbf{Concept Tokens from \emph{eGeMAPS}.} We map continuous \emph{eGeMAPS} descriptors into a set of categorical \emph{concept tokens} that can be appended to ALM prompts as structured auxiliary cues. 
The grouping follows the standard semantic definitions of the \emph{eGeMAPS} descriptors (as shown in~\cref{tab:concept_tokens}).
For each utterance, \emph{\emph{eGeMAPS}v02} LLDs are extracted with \emph{openSMILE} and summarised with a file-level mean vector. 

We then apply robust standardisation 
%(median/IQR by default) 
and compute the scalar concept scores \textit{ENERGY}, \textit{PITCH}, \textit{BRIGHTNESS}, \textit{DYNAMICS}, and \textit{FORMANTS} by averaging the corresponding standardised \emph{eGeMAPS} dimensions listed in~\cref{tab:concept_tokens}.

We discretise each scalar score into five quantile bins using dataset-level bin edges that are fixed across samples. On IEMOCAP, \textit{PITCH} and \textit{FORMANTS} use gender-specific bin edges, because adult male and female pitch and formant distributions differ substantially. Because FAU-Aibo contains pre-puberty children, we do not apply gender-specific binning on this dataset.
We additionally derive a categorical \textit{VOICE\_QUALITY} token from jitter/shimmer/HNR and harmonic-relations cues (\cref{tab:concept_tokens}). 
Finally, the six categorical labels are serialised into an ALM-friendly token string. Token construction uses \emph{no emotion labels}: bin edges are estimated from unlabelled audio, with gender metadata used only for pitch and formant calibration on IEMOCAP. Thus, the tokens provide a label-free, population re-representation of the input rather than an oracle emotion cue.

\begin{figure}[t]
  \centering
  \includegraphics[width=\linewidth]{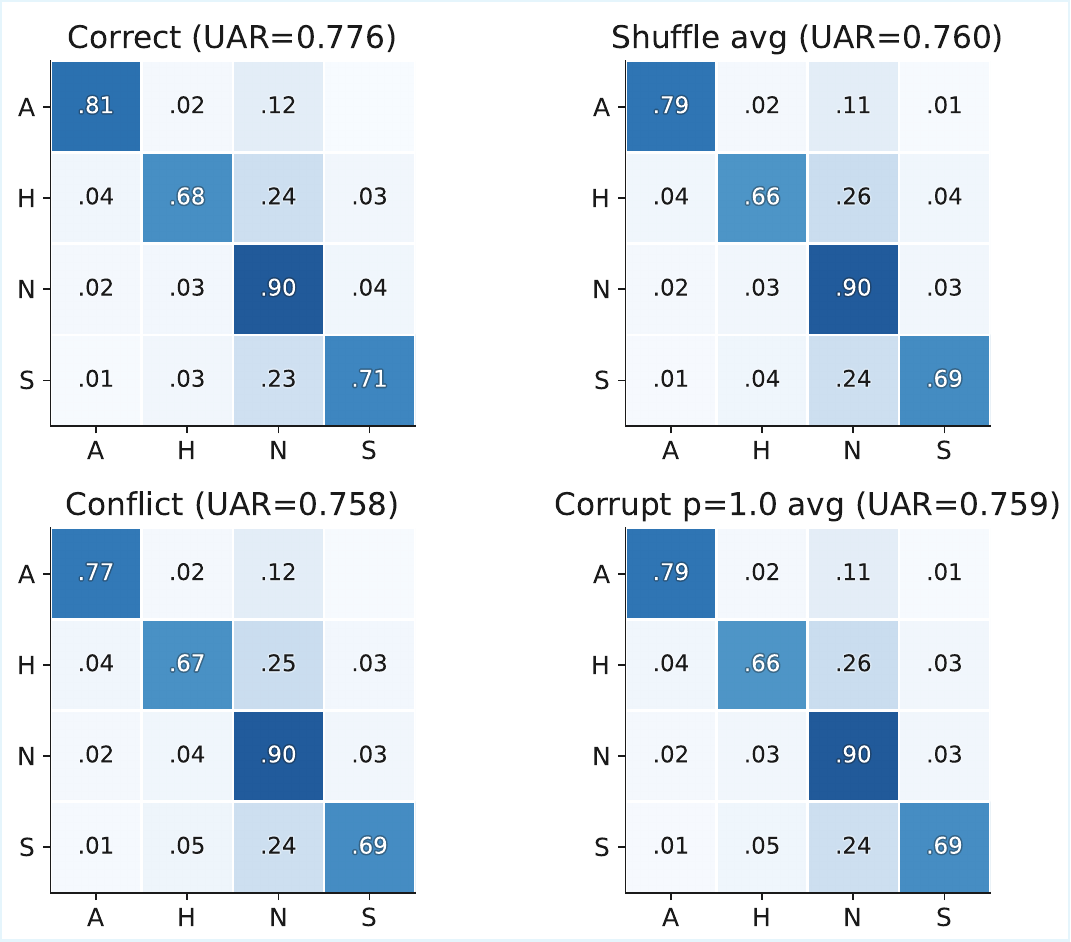}
  \caption{Row-normalised confusion matrices (rows=GT, cols=Pred) under audio reasoning interventions on AF3. Rows sum to 1; cells with value $<0.01$ are not annotated. Correct (aligned tokens) yields the best UAR, while shuffle, conflict, and full corruption reduce performance and shift errors (notably toward N).}
  \label{fig:confusion_interventions}
\end{figure}

\textbf{Audio Language Models.}
We evaluate instruction-following, autoregressive ALMs that condition an LLM on audio embeddings produced by an audio encoder. We include three open-source ALM families: Qwen2-Audio~\citep{Xu24-Q2A}, Qwen2.5-Omni \citep{Xu25-QOT}, and Audio Flamingo~3 (AF3) \citep{Goel25-AFA}, plus two task-adapted (fine-tuned) variants of Omni~\footnote{https://huggingface.co/mispeech/midashenglm-7b-0804-fp32} and AF3~\footnote{\url{https://huggingface.co/nvidia/audio-flamingo-3\#think-mode-reasoning-with-peft-adapter-af-think}}. To isolate the effect of explicit concept tokens, we use two prompt formats: \emph{audio-only} and \emph{audio+concept tokens}, where six categorical tokens are appended as structured auxiliary cues. We additionally report audio--text pretraining baselines (CLAP, ParaCLAP, SmoothCLAP) on IEMOCAP as non-generative reference points~\citep{Elizalde23-CLA,Jing24-PTA,Jing26-SSE}. Pengi is an autoregressive model~\citep{Deshmukh24-PAL}; however, since it is not designed for instruction following, prior work adapts it to zero-shot SER by mapping its generated text to the closest emotion label~\citep{Jing26-SSE}. Hence, we treat Pengi as a reference baseline and do not include it in our intervention-based tests.

\textbf{Reasoning under Token Interventions.}
We continue with \textbf{AF3} since it achieves higher token-augmented performance on both datasets: on AIBO5 test $UAR^{+}=.269$ vs $.240$ (qw-omni), and on IEMOCAP $UAR^{+}=.776$ vs $.582$, making it the strongest overall backbone for follow-up analyses.

Next, we probe audio-grounded reasoning by applying token-only interventions that modify the \emph{alignment} and \emph{quality} of the symbolic cue channel while keeping the audio fixed.

Let $x_i$ denote an utterance with ground-truth emotion $y_i$, and let $\mathbf{c}_i$ be its associated concept-token sequence (derived from \emph{eGeMAPS}). For each utterance, we construct perturbed token sequences $\tilde{\mathbf{c}}_i$ and run the same model/prompt template, measuring changes in UAR and confusion patterns.

\begin{figure}[t]
  \centering
  \includegraphics[width=\linewidth]{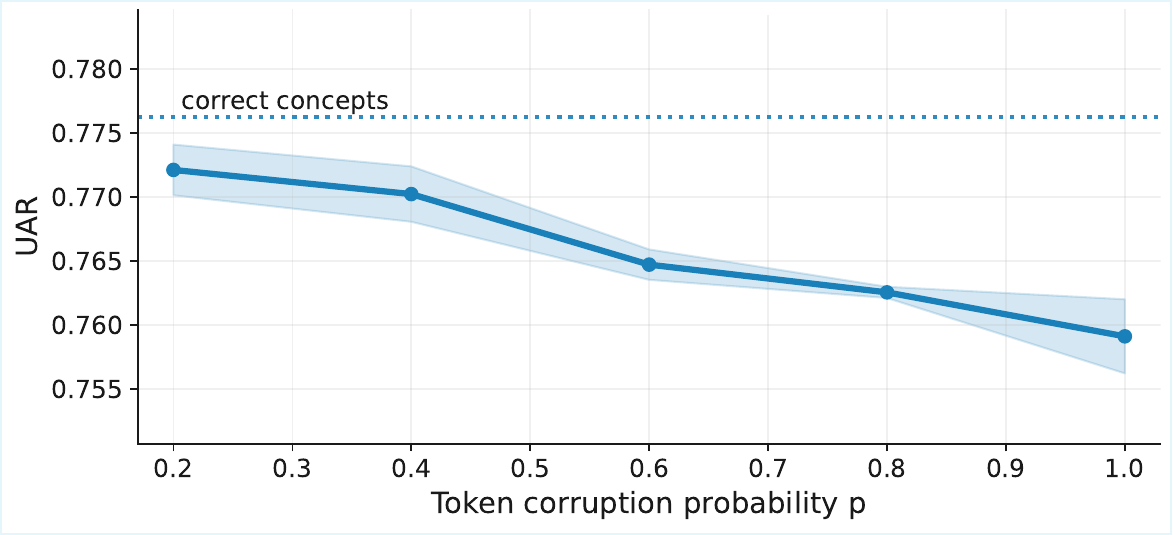}
    \caption{Token corruption curve on AF3: UAR as a function of token corruption probability $p$ (mean $\pm$ 95\% CI across seeds). Performance degrades monotonically as token quality decreases (Spearman $\rho\!=\!-.90$, $p\!<\!10^{-4}$ across the $5\times5$ $p$/seed runs), with the correct-tokens baseline shown as a reference.}
  \label{fig:corruption_curve}
\end{figure}

\textbf{Correct (aligned concept tokens).} We append the original tokens for each utterance, i.e., $\tilde{\mathbf{c}}_i=\mathbf{c}_i$. This condition represents the intended, token-aligned setting.

\textbf{Shuffle (misaligned concept tokens, distribution preserved).} We randomly reassign token sequences across utterances within the evaluation split:
$\tilde{\mathbf{c}}_i=\mathbf{c}_{\pi(i)}$ for a random permutation $\pi$. This preserves the marginal distribution of tokens but breaks the sample-level alignment between audio and concept tokens. We repeat over multiple seeds and report the mean (and confidence intervals where applicable).

\textbf{Conflict (contradictory concept tokens).} We construct tokens that systematically contradict the original ones by inverting the ordinal bins for each acoustic feature (e.g., VERY\_LOW to VERY\_HIGH, LOW to HIGH, MID to MID).
Thus, we replace each concept by its opposite category and append it as $\tilde{\mathbf{c}}_i$. This differs from Shuffle, which preserves marginal distributions across the split but breaks sample-level alignment without enforcing contradiction.

\textbf{Corrupt($p$) (graded token corruption).} To probe sensitivity to the auxiliary cues, we corrupt concept tokens with probability $p \in [0,1]$. For each token position, with probability $1-p$ we keep the original token and with probability $p$ we replace it with a randomly sampled token from the same token vocabulary. Varying $p$ yields an \emph{corruption curve}: if the model relies on the concept tokens, performance should degrade monotonically as $p$ increases.

Across all interventions, the audio input is held constant; only the appended tokens are modified. This isolates the contribution of structured cues to the model's decision process.

\section{Results}
\cref{tab:aibo5,tab:iemocap} report UAR without appended concept tokens ($UAR^{-}$) and with concept tokens ($UAR^{+}$).
On AIBO5 (dev), token augmentation causes consistent gains across all evaluated ALMs; qw-omni attains the strongest overall dev performance ($UAR^{+}=.279$), and AF3 improves reliably across splits.
On AIBO5 (test) and IEMOCAP (4-way), concept tokens again improve UAR for both qw-omni and AF3, with AF3 achieving the best overall IEMOCAP result ($UAR^{+}=.776$); on IEMOCAP, both token gains are significant ($p\!<\!10^{-5}$, with bootstrap CIs on $\Delta$ excluding zero).

The tokens themselves are class-informative in human-readable form (\cref{fig:dotstrips}); for example, sadness concentrates in the lowest \textit{ENERGY} and \textit{DYNAMICS} bins, providing the inspectable evidence channel that motivates their use.
We next analyse token sensitivity on AF3 under controlled, token-only interventions. Aligned tokens significantly outperform every perturbation, Conflict ($\Delta $UAR$ =+.018$, $95\%\,\mathrm{CI}\,[.011,.025]$, $p\!=\!4\times10^{-7}$), Shuffle, and full Corruption (both $p\!<\!10^{-4}$), and UAR degrades monotonically as corruption increases (Spearman $\rho\!=\!-.90$, $p\!<\!10^{-4}$; \cref{fig:corruption_curve}). \cref{fig:confusion_interventions} shows that perturbations modestly increase Neutral confusions relative to aligned tokens, consistent with a weakening of class-specific cues. 
These conditions decompose the aligned-token gain ($+.022$): misaligned but distribution-preserving tokens (Shuffle $.760$, Corrupt $p{=}1$: $.759$) retain only about $+.005$ over the audio-only baseline ($.754$), indicating a small \emph{format prior} from the structured block itself. The remaining gain requires \emph{correct concept content}. All perturbed conditions remain marginally above the baseline (e.g., Conflict $.758$ vs.\ $.754$, $p\!=\!.014$; Holm-corrected $p\!<\!.05$), while aligned tokens significantly outperform every perturbation. Thus, AF3 does not simply follow the symbolic cue channel, but integrates it with the audio.

The same pattern also clarifies the role of the concept tokens across datasets. On AIBO5, gains are smaller in absolute terms but consistently positive across zero-shot backbones, including the official test split. On IEMOCAP, the larger improvement for qw-omni ($+.041$) suggests that weaker audio-only prompting can benefit substantially from explicit acoustic re-representations, whereas AF3 starts from a much stronger audio-only baseline and still gains $+.022$. Thus, concept tokens are not merely compensating for a broken audio pathway; they also improve a strong ALM that already performs well from speech alone. This supports the interpretation that the tokens expose information that is present in the signal but not fully exploited by the model under the audio-only prompt.

At the same time, the intervention results caution against interpreting the tokens as an independent decision channel. If the model simply followed the symbolic cue block, contradictory tokens should have caused a much larger collapse. Instead, corrupted and conflicting prompts reduce the aligned-token advantage but remain close to the audio-only baseline. We therefore interpret the results as evidence for partial, content-sensitive use of symbolic acoustic evidence under audio grounding. The precise mechanism remains open: the model may use the tokens as soft calibration cues, as an attention-directing prompt structure, or as a weak semantic prior over acoustically plausible emotions. Distinguishing between these possibilities requires future work with internal model interventions or models trained to generate the concepts themselves.

\section{Conclusion}

We asked whether interpretable acoustic concepts can serve as intermediate evidence for ALM-based SER, testing the oracle setting in which label-free concept tokens are appended to the prompt while the audio remains unchanged. Across two corpora and multiple ALMs, aligned tokens consistently improve UAR. Token-only interventions show that most of this gain requires correct concept content: performance degrades monotonically under corruption, errors shift toward Neutral, yet predictions remain close to the audio-only baseline under misleading cues. Contemporary ALMs therefore appear to integrate symbolic acoustic evidence with the signal rather than blindly following the prompt. Conversely, the fact that partly redundant re-representations of audible properties help at all suggests a perception--utilisation gap: ALMs do not fully exploit paralinguistic information already present in their input.

These findings motivate two next steps. First, ALMs could be trained to generate concept-level intermediate steps themselves, turning the oracle setting studied here into an interpretable audio reasoning process. Second, since affective evidence is intrinsically continuous and ALMs already ingest speech as continuous embeddings, verbalised tokens could be replaced with continuous, concept-conditioned prompts in the spirit of latent-space reasoning~\cite{Hao24-TLM}. Whether symbolic or latent intermediate evidence is the better interface for audio reasoning remains an open empirical question that could be investigated in future work.

\bibliographystyle{IEEEtran}
\bibliography{mybib}

\end{document}